\documentstyle[prl,aps,epsf]{revtex}

\input{psfig}

\newcommand{\beq}{\begin{equation}}
\newcommand{\eeq}{\end{equation}}

\newcommand{\mycaption}[1]
   {\renewcommand{\baselinestretch}{1.0}\small\normalsize
    \caption{#1}
    \renewcommand{\baselinestretch}{1.5}\small\normalsize}

\newcommand{\kpzpzd}{$K_L \rightarrow 2\pi^0, 
                \pi^0 \rightarrow e^+e^-\gamma $}
\newcommand{\kpzpzD}{$K_L \rightarrow 2\pi^0_D$}

\newcommand{\kpzpzpzd}{$K_L \rightarrow 3\pi^0, 
                \pi^0 \rightarrow e^+e^-\gamma$}
\newcommand{\kpzpzpzD}{$K_L \rightarrow 3\pi^0_D$}

\newcommand{\kpzee}{$K_L \rightarrow \pi^0 e^+e^-$}
\newcommand{\kpzeeg}{$K_L \rightarrow \pi^0 e^+e^-\gamma$}

\newcommand{\keegg}{$K_L \rightarrow e^+e^-\gamma\gamma$}

\newcommand{\keeg}{$K_L \rightarrow e^+e^-\gamma$}
\newcommand{\kpzgg}{$K_L \rightarrow \pi^0\gamma\gamma$}
\newcommand{\pz}{$\pi^0$}

\newcommand{\ee}{$e^+e^-$}

\begin{document}
\title{First Observation of the decay \kpzeeg}
\author{\parindent=0.in
\parskip 0 in
A.~Alavi-Harati$^{12}$,
T.~Alexopoulos$^{12,\star}$,
M.~Arenton$^{11}$,
K.~Arisaka$^2$,
S.~Averitte$^{10}$,
A.R.~Barker$^5$,
L.~Bellantoni$^7$,
A.~Bellavance$^9$,
J.~Belz$^{10,\ddagger}$,
R.~Ben-David$^{7}$,
D.R.~Bergman$^{10}$,
E.~Blucher$^4$, 
G.J.~Bock$^7$,
C.~Bown$^4$, 
S.~Bright$^4$,
E.~Cheu$^1$,
S.~Childress$^7$,
R.~Coleman$^7$,
M.D.~Corcoran$^9$,
G.~Corti$^{11}$, 
B.~Cox$^{11}$,
M.B.~Crisler$^7$,
A.R.~Erwin$^{12}$,
R.~Ford$^7$,
A.~Glazov$^4$,
A.~Golossanov$^{11}$,
G.~Graham$^{4,\S,\dagger}$, 
J.~Graham$^4$,
K.~Hagan$^{11}$,
E.~Halkiadakis$^{10}$,
J.~Hamm$^1$,
K.~Hanagaki$^{8,\bullet}$,  
S.~Hidaka$^8$,
Y.B.~Hsiung$^7$,
V.~Jejer$^{11}$,
D.A.~Jensen$^7$,
R.~Kessler$^4$,
H.G.E.~Kobrak$^{3}$,
J.~LaDue$^5$,
A.~Lath$^{10}$,
A.~Ledovskoy$^{11}$,
P.L.~McBride$^7$,
P.~Mikelsons$^5$,
E.~Monnier$^{4,*}$,
T.~Nakaya$^{7,\parallel}$,
K.S.~Nelson$^{11}$,
H.~Nguyen$^7$,
V.~O'Dell$^7$, 
M.~Pang$^7$, 
R.~Pordes$^7$,
V.~Prasad$^4$, 
B.~Quinn$^4$,
E.J.~Ramberg$^7$, 
R.E.~Ray$^7$,
A.~Roodman$^{4,\&}$, 
M.~Sadamoto$^8$, 
S.~Schnetzer$^{10}$,
K.~Senyo$^{8,\#}$, 
P.~Shanahan$^7$,
P.S.~Shawhan$^{4,\P}$,
J.~Shields$^{11}$,
W.~Slater$^2$,
N.~Solomey$^4$,
S.V.~Somalwar$^{10}$, 
R.L.~Stone$^{10}$, 
I.~Suzuki$^{8,\S}$,
E.C.~Swallow$^{4,6}$,
S.A.~Taegar$^1$,
R.J.~Tesarek$^{10,\S}$, 
G.B.~Thomson$^{10}$,
P.A.~Toale$^5$,
A.~Tripathi$^2$,
R.~Tschirhart$^7$,
S.E.~Turner$^2$,
T.~Uchizawa$^1$, 
Y.W.~Wah$^4$,
J.~Wang$^1$,
H.B.~White$^7$, 
J.~Whitmore$^7$,
B.~Winstein$^4$, 
R.~Winston$^4$, 
T.~Yamanaka$^8$,
E.D.~Zimmerman$^{4,\pounds}$\\
\vspace*{.1 in} 
\footnotesize
$^1$ University of Arizona, Tucson, Arizona 85721 \\
$^2$ University of California at Los Angeles, Los Angeles, California 90095 \\
$^{3}$ University of California at San Diego, La Jolla, California 92093 \\
$^4$ The Enrico Fermi Institute, The University of Chicago, 
Chicago, Illinois 60637 \\
$^5$ University of Colorado, Boulder, Colorado 80309 \\
$^6$ Elmhurst College, Elmhurst, Illinois 60126 \\
$^7$ Fermi National Accelerator Laboratory, Batavia, Illinois 60510 \\
$^8$ Osaka University, Toyonaka, Osaka 560-0043 Japan \\
$^9$ Rice University, Houston, Texas 77005 \\
$^{10}$ Rutgers University, Piscataway, New Jersey 08854 \\
$^{11}$ The Department of Physics and Institute of Nuclear and 
Particle Physics, University of Virginia, 
Charlottesville, Virginia 22901 \\
$^{12}$ University of Wisconsin, Madison, Wisconsin 53706 \\
$^{\dagger}$ To whom correspondence should be addressed. \\
$^{\star}$ Current address, National Technical University, 175 73, Athens,
Greece\\
$^\ddagger$ Current address, Montana State University, Bozeman, Montana 59717\\
$^{\bullet}$ Current address, Princeton University, Princeton, 
New Jersey 08544\\
$^{*}$ Current address C.P.P. Marseille/C.N.R.S., France \\
$^{\parallel}$ Current address, Kyoto University, 606-8502 Japan\\
$^{\&}$ Current address Stanford Linear Accelerator Center, Stanford,
California 94309 \\
$^{\#}$ Current address, Nagoya University, Nagoya 464-8602 Japan \\
$^\P$ Current address, California Institute of Technology, Pasadena, 
California 91125 \\
$^\S$ Current address Fermi National Accelerator Laboratory, Batavia, 
 Illinois 60510 \\
$^{\pounds}$ Current address, Columbia University, New York, New York 10027\\
\vspace*{.1 in}
\centerline{ \bf The KTeV Collaboration}
\normalsize
}

\maketitle
\begin{abstract}

We report on the first observation of the decay
\kpzeeg\ by the KTeV E799 experiment at Fermilab. 
Based upon a sample of 48 events with an estimated background of 
$3.6 \pm 1.1$ events, we measure the \kpzeeg\ branching ratio to be
$(2.34 \pm 0.35 \pm 0.13)\times 10^{-8}$. Our data 
agree with recent ${\cal O}(p^6)$ calculations in chiral perturbation 
theory that include contributions from vector meson exchange through the 
parameter $a_V$.  A fit was made to the \kpzeeg\ data for $a_V$ with the result
$-0.67 \pm 0.21 \pm 0.12$, which is consistent with previous 
results from KTeV. 

\vspace*{0.1in}
\end{abstract}

\newpage
\narrowtext
\twocolumn

\parindent=0.3in
\normalsize


In this paper, we present the first observation of the decay mode \kpzeeg.
A previous search by the E162 experiment at KEK has established 
BR(\kpzeeg) $< 7.1\times 10^{-7}$ at the 90\% confidence level\cite{bb:murak}.
The \kpzeeg\ data presented in this paper were also used to measure 
the parameter $a_V$, which is an important ingredient in the calculation
of the branching ratio of this decay mode 
using Chiral Perturbation Theory (CHPT).  
CHPT is a low energy effective theory of 
QCD which has been a very useful tool for describing kaon decays in which 
long distance effects are expected to dominate.  
An ${\cal O}(p^6)$ calculation using CHPT by Donoghue and 
Gabbiani\cite{bb:donoghue2} predicts a 
branching ratio for \kpzeeg\ of $2.4\times 10^{-8}$,
contrasted with $1.0\times 10^{-8}$ from the ${\cal O}(p^4)$ calculation.
Our measurement can distinguish between these two predictions.

The rare decay \kpzeeg\ is intimately related to the decay \kpzgg\ via the
internal conversion of one of the photons into an \ee\ pair.
The ${\cal O}(p^4)$ calculation of \kpzgg\
requires no free parameters, but
the estimated branching ratio was found
to be a factor of three lower than the measured value \cite{bb:ecker_orig}.
The CHPT calculation extended to ${\cal O}(p^6)$ was 
found to be able to reproduce the branching ratio as well as the distinctive 
$m_{\gamma\gamma}$ spectrum\cite{bb:dambrosio}.  However,  the cost was to  
introduce a free parameter into the theory, $a_V$,  which parameterizes the 
contribution of vector meson exchange to the decay amplitude.   
The parameter $a_V$ was estimated to be -0.96 in 
Reference \cite{bb:donoghue} from fits to 2$\pi$ and 
3$\pi$ decay data in the kaon system.  
A recent measurement of $a_V$ by the KTeV experiment in the \kpzgg\ system 
finds $a_V$ to be $-0.72\pm0.05\pm0.06$, 
and has confirmed that the shape of the 
$m_{\gamma\gamma}$ spectrum is reproduced by the theory\cite{bb:cheu}.

The decay \kpzeeg\ is also interesting because of its relationship
to the CP violating decay \kpzee. Since a 
significant fraction of the total \kpzee\ amplitude 
proceeds through a CP conserving two photon intermediate state, 
$K_L\rightarrow \pi^0 \gamma^* \gamma^*$, 
it is essential to probe the \kpzeeg\ decay dynamics 
in order to better disentangle the CP conserving
amplitudes from the CP violating ones in \kpzee\cite{bb:donoghue}.
Also, the rate of \kpzeeg\ is expected to be several orders of 
magnitude higher than the rate of the CP violating rare decay \kpzee.  
In the soft photon region, it can be a source of 
background for the search for \kpzee.  
  
%
%
The data analyzed here were collected during the 1997 
runs of E799.  The KTeV detector itself is described in detail elsewhere 
\cite{bb:ggraham}. 
The principal KTeV detectors used in this analysis include a pure
CsI electromagnetic calorimeter, a nearly hermetic 
lead-scintillator photon veto system, and a 
charged particle spectrometer system.  

The CsI calorimeter \cite{bb:roodman} 
is composed of 3100 blocks in a 1.9 m by 1.9 m array
that is 27 radiation lengths deep. Two 15 cm by 15 cm holes are located
near the center of the array for the passage of two neutral kaon beams.
For electrons with energies between 5 GeV and 60 GeV,
the calorimeter energy resolution was better than 
1\%. The position resolution
for electromagnetic clusters in the calorimeter is approximately 1 mm. 
\begin{figure}
\centering
\mbox{\epsfxsize 8.5cm 
       \epsfbox{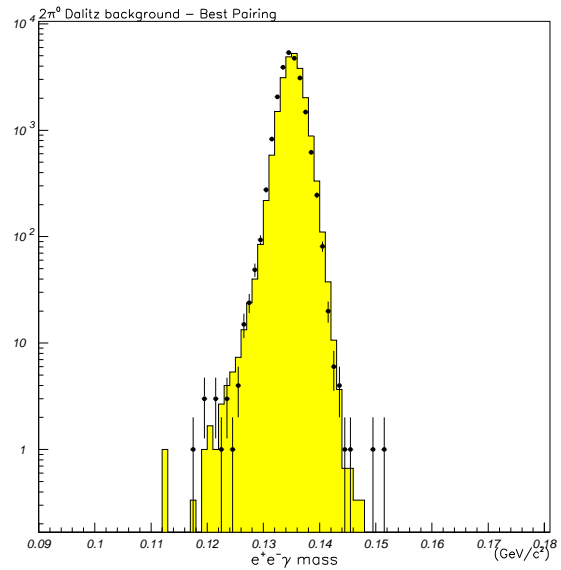}}
 \caption{The $m_{ee\gamma}$ distribution in E799 data (points) 
 is compared with \kpzpzD\  Monte Carlo (histogram) including
all cuts described in this analysis but for the
cut against correctly paired \kpzpzD\ background.    
A small shift in the $\pi^0$ mass from calibration effects 
does not significantly contribute to systematic error 
in this measurement.}
\label{fig:meeg}
\end{figure}

The charged particle spectrometer 
comprises four drift chambers. Two are located upstream and two downstream 
of an analyzing magnet with
a transverse momentum kick of 0.205 GeV/$c$.  Each drift chamber has four 
drift planes, with two planes oriented along the X direction and 
two along the Y 
direction transverse to the beams. Each drift plane is staggered by 0.5 cell 
with respect to its partner and was better than 99\% 
efficient for recording hits from passing tracks.  
The momentum resolution 
was $ 0.38\% + 0.016\% \times p(GeV)$.  

The fiducial decay volume is a vacuum
tank of approximately 60 meters length located immediately upstream 
of the first drift chamber.  The vacuum was maintained
at better than $10^{-6}$ Torr for the data considered here.
The charged particle spectrometer and vacuum tank are
surrounded by 9 detectors arrayed along the length of the detector that 
veto stray photons.


The \kpzeeg\ final state consists of three photons, two of which proceed 
from \pz\ decay, and two electrons.  \kpzeeg\ events are recorded if they
satisfy the following trigger requirements. There must be at least 2 hits in 
one of two scintillator hodoscopes and at least 1 hit in the other.
In addition, there must be at 
least one hit in one of the two upstream drift chambers.  The event must
deposit more than approximately 25 GeV of total energy in the CsI 
calorimeter and deposit little energy in the photon 
vetoes.  The event is vetoed if energy is deposited in a scintillator
hadron veto located downstream of the CsI and a 10cm thick lead wall.  
The trigger includes a hardware cluster processor that counts the number of 
calorimeter clusters of contiguous blocks of CsI  
with energies above 1 GeV \cite{bb:hcc}. The total number of electromagnetic 
clusters at the trigger level in the CsI calorimeter is required to be  
greater than or equal to four. These trigger 
requirements also select \kpzpzd\
(\kpzpzD) events that we use as normalization for the \kpzeeg\ signal.

In the offline analysis, we require exactly five reconstructed electromagnetic 
clusters in the CsI calorimeter each with
energy greater than 2.0 GeV. Two of the clusters must 
match the extrapolated positions of the downstream components of the charged 
tracks.  
In order to identify the two charged tracks as electrons, the ratio of energy
of the matched cluster as measured by the CsI (E) to track momentum as 
measured 
by the spectrometer (p) must lie in the interval $0.95 < E/p < 1.05$. 

We calculate the kaon decay vertex by 
extrapolating track segments upstream of the analysis magnet back
towards the target. 
The vertex position is then used to reconstruct a
\pz\ from each of the three possible $\gamma\gamma$ pairings.  We 
rank the pairings by closeness of reconstructed mass to the 
nominal \pz\ mass and designate
the best $\gamma\gamma$ pair to be the reconstructed \pz.
We also require that the best pairing choice satisfies 
$|m_{\gamma\gamma} - m_{\pi^0}|< 5.0 $ MeV/$c^2$.
In the normalization 
mode analysis, an additional requirement is made on the invariant mass of 
the $e^+e^-\gamma$ system;
 $|m_{ee\gamma} - m_{\pi^0}|< 6.0 $ MeV/$c^2$.  
Events with missing momentum are  supressed by a cut on the square of the 
transverse momentum of the reconstructed kaon
defined with respect to the line connecting the kaon decay vertex and the 
target ($p_T^2$).  We require $p_T^2 < 300 (MeV/c)^2$.  The 
decay vertex is required to be well within the vacuum decay volume.  
We require $98 m < Z_{vtx} < 157 m$ from target, and the
reconstructed kaon momentum is required to be between 30 and 210 GeV/c.

\kpzpzD\  and \kpzpzpzd\ (\kpzpzpzD) decays are 
a source of background when photons are lost and/or mispaired. 
Backgrounds resulting from kaon 
decays with extra soft charged particles in the final state, such 
as 3\pz\ decays where two of the \pz\ undergo Dalitz decay, are easily 
removed by discarding events with extra in-time hits in the  
region of the spectrometer upstream of the analysis magnet.  
Such events may fake two track events  if 
soft tracks are swept out of the detector by the analysis magnet.

The \kpzpzD\ events are well simulated by our Monte Carlo as shown in 
Figure~\ref{fig:meeg}.  The mass peaks agree to 
within 0.15\% \cite{bb:ggraham}.   
In order to remove \kpzpzD\ decays in the signal mode analysis we require
the mass $m_{e^+e^-\gamma}$ not lie between 0.115 GeV/$c^2$ and 
0.150 GeV/$c^2$.  
Mispaired
\kpzpzD\ events, where the incorrect pairing of photons is chosen, 
constitute a possible background. 
We remove mispaired \kpzpzD\
events by considering each of the other possible photon 
pairings and discarding events which have reconstructed values of 
$m_{\gamma\gamma}$  and $m_{ee\gamma}$  that are 
consistent with \kpzpzD\ decay.  
Specifically, events are discarded if 
$\sqrt{(m_{\gamma\gamma}-m_{\pi^0})^2+(m_{ee\gamma}-m_{\pi^0})^2} < $ 8 MeV or 
6 MeV in the case of the second best or worst pairing choice respectively.
 
\begin{figure}
\centering
\mbox{\epsfxsize 8.5cm 
       \epsfbox{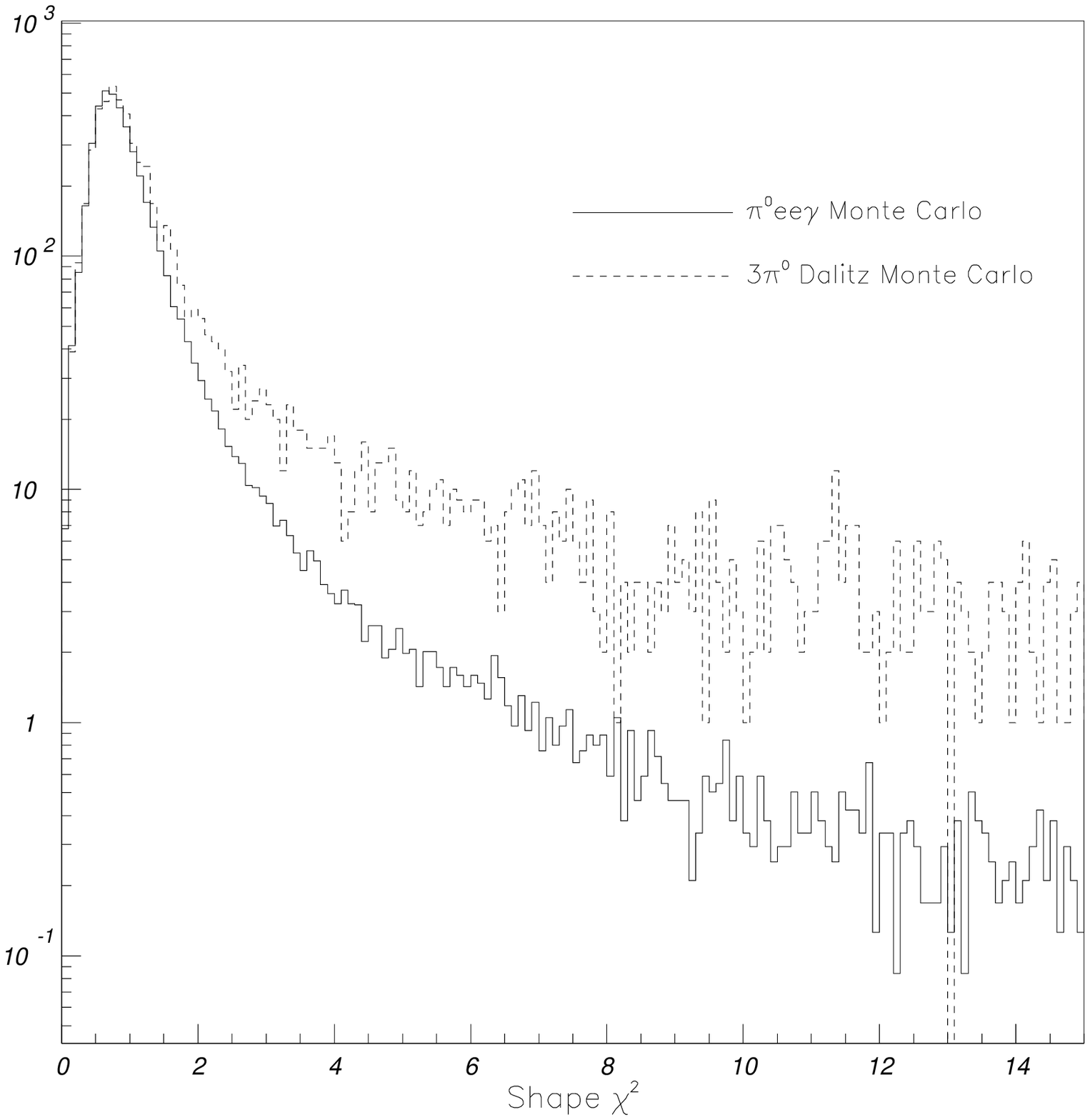}}
\caption{The calorimeter shape $\chi^2$ distribution for \kpzeeg\ and 
\kpzpzpzD\ Monte Carlo samples.
The larger tail in the \kpzpzpzD\ sample is indicative of 
a higher proportion of fusions.}
\label{fg:shape}
\end{figure}

\kpzpzpzD\ decays are the most difficult source of background
to suppress in this analysis.  These events can contribute to the 
background if two photons escaped detection or are
fused in the calorimeter.  In order to reduce backgrounds from 
\kpzpzpzD\ events with 
lost photons, we require less than 0.5 MIP activity in each of the
photon veto detectors upstream of the spectrometer
in the offline analysis.  
Some remaining \kpzpzpzD\ background 
results from
events in which two or more photons fuse together in the
CsI calorimeter.   In order to reduce this background,  a shower shape
$\chi^2$ was calculated for the distribution of energy deposited in the 
central crystals comprising the energy cluster compared to 
distribution of energy for a single photon.
Figure~\ref{fg:shape} shows the distribution 
of shape $\chi^2$ for both \kpzpzD\ and \kpzpzpzD\ Monte Carlo. 
By requiring the maximum $\chi^2$ to be less 
than 5.0, we are able to effectively remove background due to \kpzpzpzD\ 
fusions.  

\begin{figure}
\centering
\mbox{\epsfxsize 8.5cm 
       \epsfbox{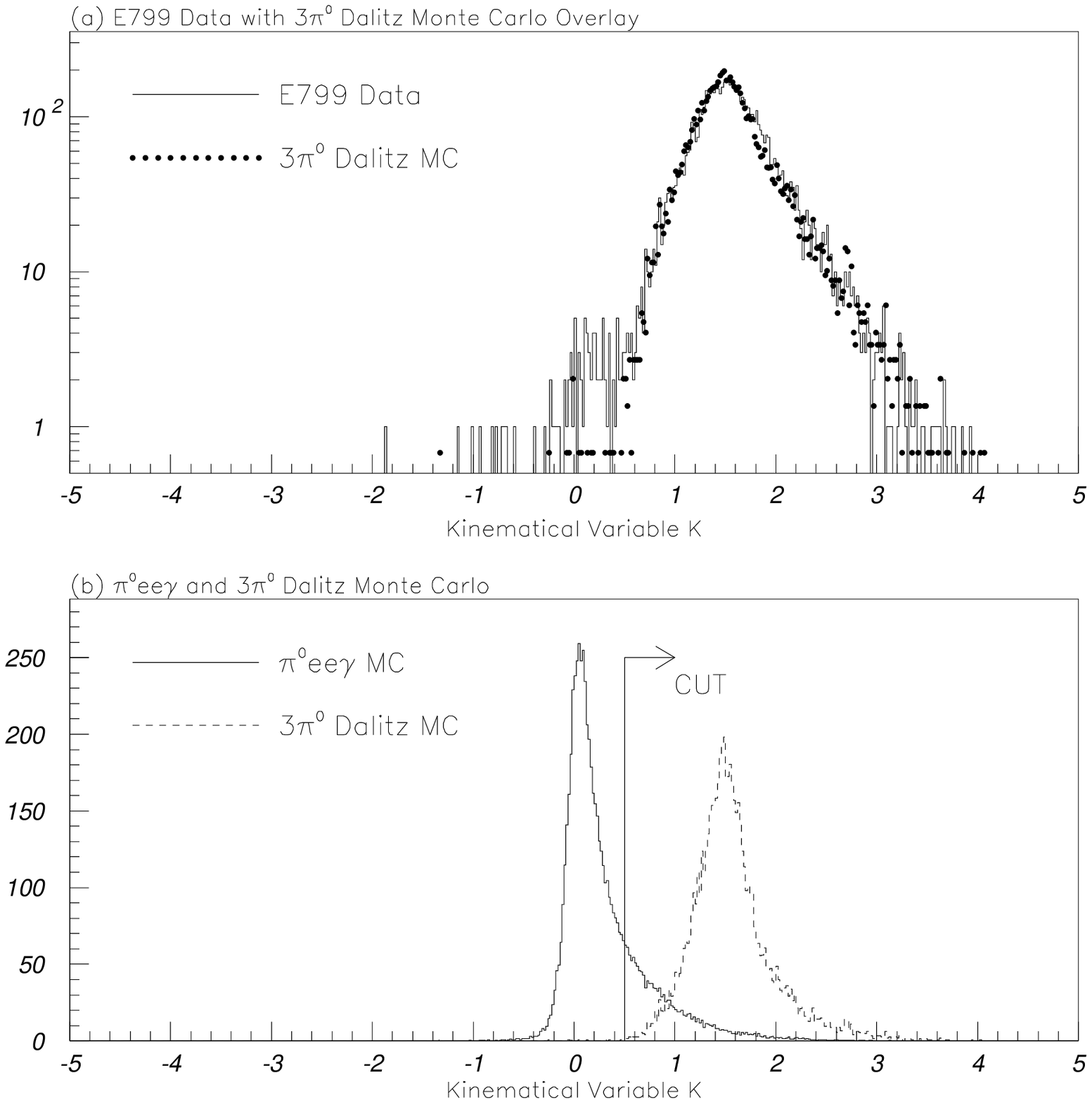}}
\caption{The distribution of the kinematical variable described in the text,
 $\kappa$,  for \kpzpzpzD\ and
\kpzeeg\ Monte Carlo samples. The upper plot shows \kpzpzpzD\ in the E799 
data overlaid with Monte Carlo.  The lower plot shows \kpzpzpzD\ Monte 
Carlo with \kpzeeg\ Monte Carlo.  }
\label{fg:kine}
\end{figure}

A cut is also made on the kinematics of the \kpzpzpzD\ decay.  
The kinematical variables 
used are the $3\gamma$ invariant 
mass, $m_{3\gamma}$, and the square of the longitudinal momentum 
of the $e^+e^-$ system in the kaon rest frame, 
${\cal{P}}^2$.  
In order to make a cut in the ${\cal{P}}^2$-$m_{3\gamma}$ plane, 
a single variable $\kappa$ was constructed from these two variables. 
The parameter 
$\kappa = \log{{\cal{P}}^2} - 15.6\times m_{3\gamma} + 4.6$, where the offset 
of 4.6 is chosen so that the the \kpzeeg\ Monte Carlo 
distribution in $\kappa$ has its 
maximum at $\kappa=0$.   Figure \ref{fg:kine} shows the
distribution in $\kappa$ for \kpzpzpzD\ in data and Monte Carlo and for
\kpzeeg\ in Monte Carlo.   
The parameter 
$\kappa$ is thus a convenient single variable kinematical discriminant 
between \kpzeeg\ signal and \kpzpzpzD\ background \cite{bb:ggraham}. 
In order to remove the remaining 
\kpzpzpzD\ background,  we require $\kappa < 0.5$.

\begin{figure}
\centering
\mbox{\epsfxsize 8.5cm 
       \epsfbox{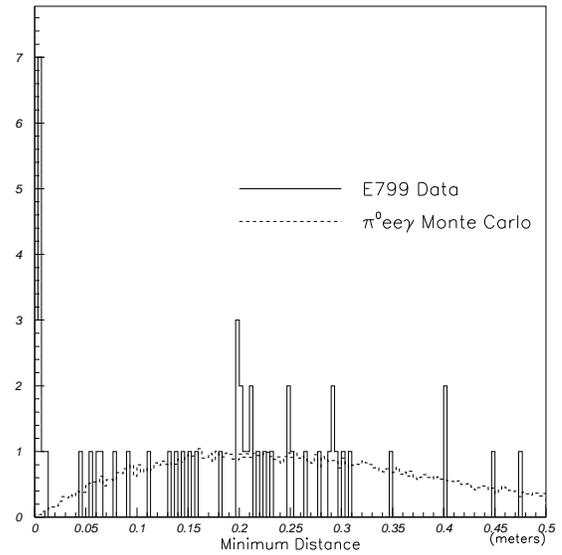}}
\caption{The minimum distance between the extrapolated position of 
electrons at the CsI face using the tracks segments upstream of the 
analysis magnet and unmatched (photon) clusters is 
shown for the E799 data and for \kpzeeg\ Monte Carlo. }
\label{fig:dmin}
\end{figure}

  A more subtle background can come from \keeg\ events accompanied by 
one or two bremsstrahlung photons plus accidental energy in-time with the 
event.  In order to remove such 
events, a quantity $D_{min}$ is calculated. The parameter 
$D_{min}$ is the minimum distance
between the position of an electron  
at the CsI extrapolated using the track segment upstream of the
analysis magnet and the position of a photon
at the CsI.  The analysis requires $D_{min}>1.25$ cm.  
This cut removes remaining backgrounds from brehmsstralung and is 99.5\%
efficient for keeping the \kpzeeg\ signal.  
Figure~\ref{fig:dmin} 
shows the distribution of $D_{min}$ for the data and 
signal Monte Carlo.  Nine events are removed from the \kpzeeg\ sample 
by this cut at this stage.  

\begin{figure}
\centering
\mbox{\epsfxsize 8.5cm 
       \epsfbox{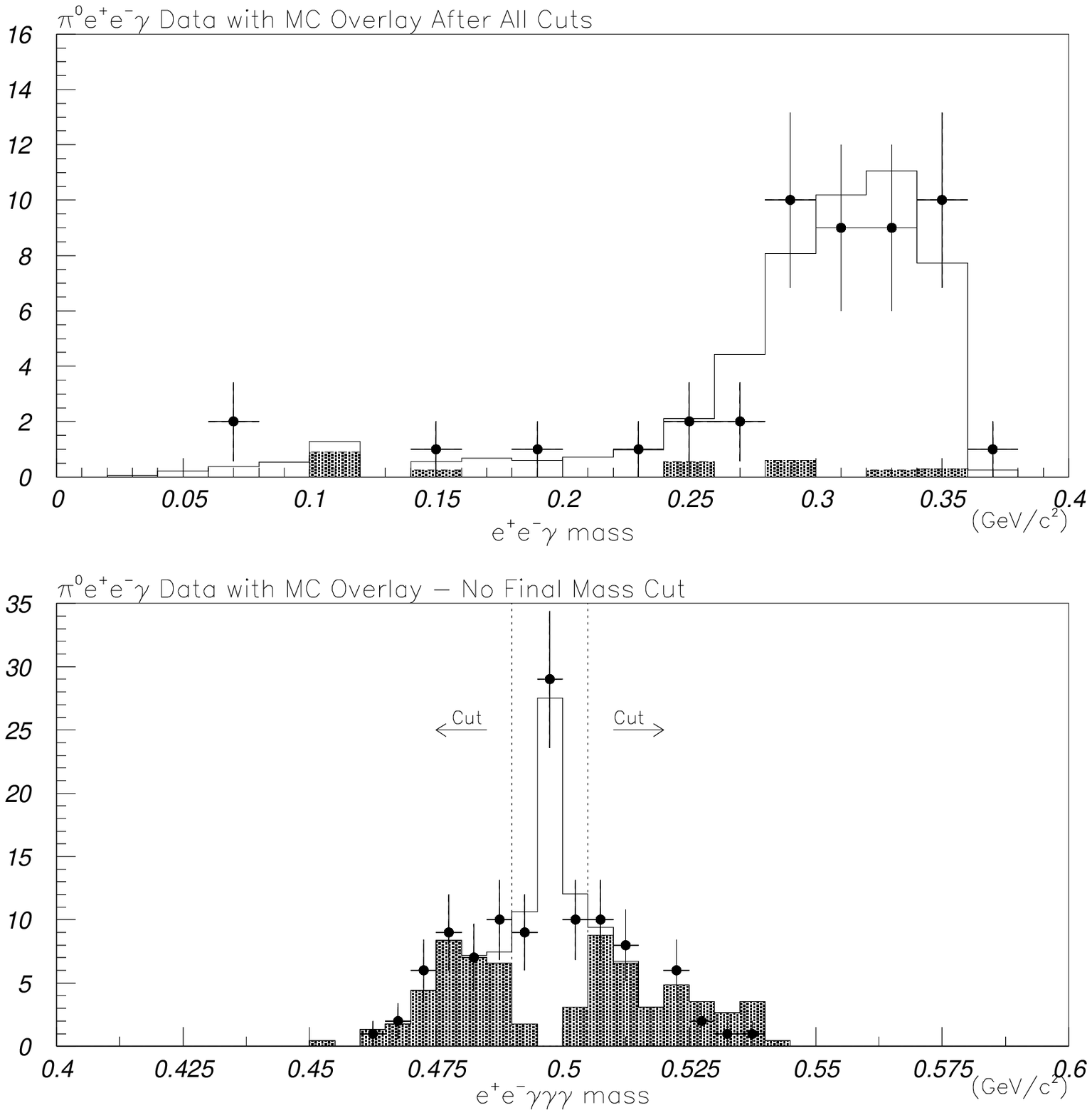}}
  \mycaption{The top plot shows the $m_{ee\gamma}$ distribution for the
final \kpzeeg\ sample.  
The $m_{ee\gamma\gamma\gamma}$ distribution in E799 data 
(points) after all cuts except the final mass cut is shown below 
compared with signal+background Monte Carlo (histogram.)  
The solid histograms are from Monte Carlo, and the points with error bars are
from the data. Background Monte Carlo alone appears in the shaded areas.}
\label{fg:meeggg}
\end{figure}

The $m_{ee\gamma}$ distribution of the final event sample after 
making all cuts is shown at top in  Figure~\ref{fg:meeggg}, and 
the $m_{ee\gamma\gamma\gamma}$ distribution is shown below without 
the final mass cut.   
In order to select the final \kpzeeg\ event sample, a cut is made on the
final state mass requiring that $m_{ee\gamma\gamma\gamma}$ lie
between 490 MeV/$c^2$ and 505 MeV/$c^2$.  We find a total of 48
candidate events. The estimated background in this sample from Monte Carlo 
include $1.6\pm0.6$ events from \kpzpzD, $1.5\pm0.9$ events from \kpzpzpzD, 
and $0.5\pm0.1$ events from \keegg\ for a total of $3.6\pm1.1$
background events.  The remaining background in the sidebands on the 
lower plot is due primarily to \kpzpzD\ decays that pass the cuts 
designed to reject Dalitz \pz\ in the final state.

 In order to determine the \kpzeeg\ branching ratio, we normalize
the \kpzeeg\ event sample to the \kpzpzD\ event sample.  
The \kpzpzD\ events are 
selected by removing the requirements against \kpzpzD\ and requiring
that $m_{e^+e^-\gamma}$ in the best pairing hypothesis 
lie within the window 0.129 GeV/$c^2$ to 
0.141 GeV/$c^2$ while keeping all other 
cuts in place.  We find 56467 \kpzpzD\ events in the E799 dataset
with negligible background, corresponding to a kaon 
flux of $2.64\times 10^{11}$ kaons decaying in the E799 fiducial volume
during the run.

The systematic uncertainty in this measurement is reduced since
we are only sensitive to differences in the relative acceptance
of the signal and normalization modes and not the absolute acceptance. 
The acceptances for the modes \kpzeeg\ and \kpzpzD\ are 
0.72\% and 0.95\%, respectively.  

 The principal systematic uncertainties in this measurement come from 
the \kpzpzD\ branching ratio (3.4\%), 
the 3\pz\ background MC statistics (3.0\%), 
varying the $a_V$ parameter in the \kpzeeg\ Monte Carlo (2.0\%), 
varying the photon shape $\chi^2$ cut (1.8\%), 
varying the kinematic cut against \kpzpzpzD\ (0.6\%), 
varying the cuts on the photon vetoes (0.5\%), 
aperture effects (0.4\%), 
varying the mass cuts (0.2\%). 
The systematic uncertainties are added in quadrature, resulting in a
total systematic uncertainty of 5.3\%.
We find the branching ratio to be 
BR(\kpzeeg)
= $(2.34 \pm 0.35(stat.) \pm 0.13(syst.))\times 10^{-8}$.  
This is consistent with the ${\cal O}(p^6)$ prediction
and not with the ${\cal O}(p^4)$ CHPT prediction for the branching ratio.
This measurement of \kpzeeg\ branching ratio also constitutes the 
first observation of this decay mode. 

  The parameter $a_V$ was determined by maximum likelihood fit of the 
data to the Monte Carlo distributions in   
$Z = \frac{m_{ee\gamma}^2}{m_K^2}$, $Q = \frac{m_{ee}^2}{m_K^2}$, 
and $Y = \frac{E_{\gamma}-E_{ee}}{m_K}$, where $E_{\gamma}$ and $E_{ee}$ 
are the energies of the photon and the $e^+e^-$ pair
in the kaon CM frame respectively.  
The fitting method
is described in more detail in \cite{bb:cheu}.   The systematic error 
in this measurement was estimated by 
varying the amount of background in the data plot by an \mbox{amount} consistent 
with the statistical error in the background and by
varying the analysis requirements.   We measure $a_V$ to  
be $ -0.67 \pm 0.21(stat.) \pm 0.12(sys.)$.  This result is consistent 
with the previous measurement by KTeV in the \kpzgg\ system.

We gratefully acknowledge the support and effort of the Fermilab
staff and the technical staffs of the participating institutions for
their vital contributions.  We also acknowledge 
G.~D'Ambrosio, F.~Gabbiani and J.~Donoghue for thoughtful discussions.
This work was supported in part by the U.S. 
Department of Energy, The National Science Foundation and The Ministry of
Education and Science of Japan.  
In addition, A.R.B., E.B. and S.V.S. 
acknowledge support from the NYI program of the NSF; A.R.B. and E.B. from 
the Alfred P. Sloan Foundation; E.B. from the OJI program of the DOE; 
K.H., T.N. and M.S. from the Japan Society for the Promotion of
Science.  P.S.S. acknowledges receipt of a Grainger Fellowship.

\end{document}